\documentclass[conference]{IEEEtran}
\IEEEoverridecommandlockouts
\usepackage[noend]{algpseudocode}
\usepackage{amsmath,amssymb,amsfonts}
\usepackage[labelformat=simple]{subcaption}

\usepackage[hidelinks]{hyperref}
\usepackage[utf8]{inputenc}
\usepackage{dblfloatfix}
\usepackage{tcolorbox}
\usepackage{algorithm}
\usepackage{listings}
\usepackage{graphicx}
\usepackage{booktabs}
\usepackage{textcomp}
\usepackage{graphicx}
\usepackage{cleveref}
\usepackage{multirow}
\usepackage{textcomp}
\usepackage{balance}
\usepackage{comment}
\usepackage{siunitx}
\usepackage{diagbox}
\usepackage{xcolor}
\usepackage{pifont}
\usepackage{bm}
\usepackage{diagbox}
\usepackage{setspace} 
\newcommand{\cmark}{\checkmark}
\newcommand{\xmark}{\ding{55}}

\usepackage[
backend=biber
,style=ieee
,minbibnames=1
,maxbibnames=99
,maxcitenames=1
,mincitenames=1
,eprint=false
,doi=false
,isbn=false
,url=false
]{biblatex}
\bibliography{ref} 
\AtBeginBibliography{\footnotesize}

\def\BibTeX{{\rm B\kern-.05em{\sc i\kern-.025em b}\kern-.08em
    T\kern-.1667em\lower.7ex\hbox{E}\kern-.125emX}}

\usepackage{tikz}
\tikzset{%
pics/mycirc/.style args={#1}{
      code = {
\node [draw, circle, fill, scale=0.5] {\color{white}#1};
}}}

\usepackage{xcolor}

\newcommand{\ad}[1]{\textcolor{blue}{#1}}
\renewcommand{\ad}[1]{#1}
\begin{document}

\title{Exploring a Test Data-Driven Method for Selecting and Constraining Metamorphic Relations}

\author{
% \vspace{1.5cm}
\IEEEauthorblockN{
	Alejandra Duque-Torres\IEEEauthorrefmark{2}, Dietmar Pfahl\IEEEauthorrefmark{2}, Claus Klammer\IEEEauthorrefmark{3}, and
    Stefan Fischer\IEEEauthorrefmark{3}}
			
	\IEEEauthorblockA{%
	\IEEEauthorrefmark{2}\textit{Institute of Computer Science  }, \textit{University of Tartu}, Tartu, Estonia \\
	E-mail: \{duquet, dietmar.pfahl\}@ut.ee}	
	\IEEEauthorblockA{
	\IEEEauthorrefmark{3}\textit{Software Competence Center Hagenberg (SCCH) GmbH}, Hagenberg, Austria \\
	E-mail:  \{claus.klammer, stefan.fischer\}@scch.at}
}

\maketitle

\begin{abstract}
Identifying and selecting high-quality Metamorphic Relations (MRs) is a challenge in Metamorphic Testing (MT). While some techniques for automatically selecting MRs have been proposed, they are either domain-specific or rely on strict assumptions about the applicability of a pre-defined MRs. This paper presents a preliminary evaluation of MetaTrimmer, a method for selecting and constraining MRs based on test data. MetaTrimmer comprises three steps: generating random test data inputs for the SUT (Step 1), performing test data transformations and logging MR violations (Step 2), and conducting manual inspections to derive constraints (Step 3). The novelty of MetaTrimmer is its avoidance of complex prediction models that require labeled datasets regarding the applicability of MRs. Moreover, MetaTrimmer facilitates the seamless integration of MT with advanced fuzzing for test data generation. In a preliminary evaluation, MetaTrimmer shows the potential to overcome existing limitations and enhance MR effectiveness.

\end{abstract}


\section{Introduction}
\label{sec:introduction}
Metamorphic Testing (MT) is a software testing approach introduced by \citeauthor{chen2020metamorphic} to address the test oracle problem. The oracle problem arises when the System Under Test (SUT) lacks an oracle or when creating one to determine whether the SUT outputs are correct is practically impossible. Unlike traditional testing approaches, MT examines the relations between input-output pairs of consecutive SUT executions, called Metamorphic Relations (MRs). MRs define how the outputs should vary in response to specific changes in the input, enabling testers to indirectly test the SUT by checking whether the inputs and outputs satisfy the MRs \cite{SANER-VST2023}. A violated MR suggests a high probability that the SUT has a fault, but no violations do not ensure a fault-free SUT. Selecting appropriate MRs is non-trivial and requires a deep understanding of the SUT and its domain. Hence, MR selection remains a significant challenge in MT. Although some techniques have been proposed for automatic MR selection, they either focus on specific domains or operate under strict assumptions about the applicability of pre-defined MRs. For instance, Predicting Metamorphic Relations (PMR) uses Machine Learning (ML) techniques to automatically select probable MRs of program methods based on extracted features from the method's Control-Flow Graph (CFG) \cite{PMR1}. The idea behind PMR is to create a model that predicts whether a specific MR can be used to test a method in a newly developed SUT. Initial PMR evaluations used path- and node-based features extracted from CFGs of Java methods and three predefined MRs to train SVM and decision tree models \cite{PMR1}. Later, the approach was extended to six predefined MRs using graph similarity measures, such as RWK and GK \cite{PMR3}. Other works have followed the PMR approach, such as \citeauthor{PMR4} who used semi-supervised learning on CFG-based features \cite{PMR4}, \citeauthor{PMR5} who applied PMR to matrix-based programs \cite{PMR5}, and \citeauthor{vst2022Aleja} who used software metrics instead of CFG-based features \cite{vst2022Aleja}.

Despite promising results from the PMR study and subsequent works, the PMR approach has significant limitations. Firstly, it relies on binary classifiers that require labelled datasets to provide examples for learning. Labelled datasets may not always be available, and obtaining them can be time-consuming. Secondly, the feature extraction process for model training is based on CFG or source code metrics, which may not account for refactoring. This limitation can affect the accuracy of the PMR approach, as refactoring can change the structure of the code and, consequently, the way MRs apply. Lastly, the binary output of PMR may not consider test data and its impact on MR applicability. This limitation implies that PMR does not consider the possibility that an MR may apply to some test data with specific characteristics and not others, leading to false positives or false negatives in MR selection.

% To address the limitations of PMR, this paper presents preliminary results of a test data-driven method for selecting and constraining MRs. Unlike PMR, this method does not rely on labelled datasets. The method involves generating random test data using a fuzzer, applying MRs to transform the data, and recording information about inputs, outputs, and MR violations during execution. Two scenarios were evaluated to assess the method's effectiveness in selecting and constraining MRs.

% In the first scenario, we select MRs based on test data while knowing the restrictions of the SUT input data. The second scenario focused on using test data to derive constraints for MRs without knowing any restrictions. The results of the evaluation were compared against those of PMR. The proposed method outperformed PMR in both scenarios, demonstrating its effectiveness in selecting and constraining MRs based on test data. The main advantage of the proposed method is its ability to identify corner cases that may be overlooked, leading to a more thorough and effective test suite. We conclude that the proposed method presents a promising alternative to the PMR approach, offering an effective way to select and constrain MRs based on test data. Its potential benefits include increased test coverage and improved testing efficiency.

Motivated by the limitations of PMR and the challenges in selecting appropriate MRs, this paper proposes \ad{MetaTrimmer}, a test data-driven method for selecting and constraining MRs and presents preliminary results. Similar to PMR, we assume a pre-defined list of MRs is available. However, \ad{MetaTrimmer} does not rely on labelled datasets and takes into account that an MR may only be applicable to test data with specific characteristics. The main goal of \ad{MetaTrimmer} is to simplify the process of selecting MRs from the pre-defined set and provide constraints to the MRs based on the test data. 

\ad{MetaTrimmer comprises three main steps: (1) Test Data Generation (TD Generation), (2) MT Process, and (3) MR Analysis. Step 1, TD Generation, is responsible for generating random test data for the SUT. In step 2, the MT Process carries out necessary test data transformations based on the MRs, and generates logs to record information about inputs, outputs, and any MR violations during the execution of the test data and the transformed test data against the SUT. MR Analysis, step 3, conducting manual inspections of violation and non-violation results and identifying specific test data or ranges where the MR is applicable to derive constraints.}
% \ad{MetaTrimmer} comprises three main modules: the Test Data Generation (TDG) module, which generates random test data using a fuzzer; the MT module, which follows the MT workflow by applying the specified change by each MR to the test data to transform it, and then verifies that the changes between the observed output with the test data and the transformed data match the changes specified by the MR after the test execution. From these two modules, we record in a Log the test data generated in the first module, the transformed data, and whether the MR is violated or not for all predefined MRs. The last module, the Analyser module, analyses the data contained in the Log to extract the constraints.
To evaluate the feasibility and effectiveness of our approach, this paper aims to answer two research questions. 

\textbf{RQ$_{1}$: [Selection] How well does \ad{MetaTrimmer} perform compared to the PMR approach in terms of MR selection?} This research question investigates the feasibility of using test data to determine the applicable MRs for a given method. Our hypothesis is that an MR that is violated 100\% of the time does not apply to the tested method, while an MR that is not violated in 100\% of the cases aligns with the tested method. To this end, we used a set of 25 methods, each labelled with a set of six predefined MRs used in the PMR evaluation by \citeauthor{PMR3}~\cite{PMR3}. We carefully designed the test data to avoid triggering any potential exceptions that the SUT may throw due to invalid inputs, as we assume that the authors of the original PMR approach also did the same. Specifically, we restrict the test data to positive integral numbers.

\textbf{RQ$_{2}$: [Constraints] How can test data be used to obtain constraints for MRs?} This research question investigates how test data can be used to generate constraints for MRs, particularly in scenarios where violations and non-violations do not reach 100\%, referred to as mixed cases. The hypothesis is that mixed cases can provide valuable insights into the behaviour of MRs and aid in generating constraints. Building upon RQ$_{1}$, where test data was limited to positive integral numbers, this question incorporates test data that can potentially cause the SUT to throw exceptions to explore how constraints for MRs can be obtained when valid input data knowledge is not available. To this end, we use the same MRs and the same 25 methods used in RQ$_{1}$.

Our preliminary results indicate that \ad{MetaTrimmer} is feasible and can efficiently select and constrain MRs based on test data. However, further empirical evaluation is required to verify the effectiveness and generalisability of \ad{MetaTrimmer}. We intend to perform additional experiments and assessments to validate \ad{MetaTrimmer} and determine any potential shortcomings and areas for improvement. Despite this, we believe that our approach can significantly contribute to the field of MRs selection.

The rest of the paper is structured as follows. \Cref{sec:background} presents the main concepts used in our study. In \Cref{sec:methodology}, we introduce \ad{MetaTrimmer}, along with the research questions, the MRs, and the SUTs used for the evaluation. In \Cref{sec:results and discussion}, we present the answers to the research questions and discuss our results. We provide the related work in \Cref{sec:related work}. Finally, we conclude the paper in \Cref{sec:conclusion}.

% \vspace{-0.5ex}

\section{{Background}}
\label{sec:background}
This section presents the key concepts used in our research. \Cref{subsec:Metamorphic Testing} introduces the MT approach. \Cref{subsec:TDG} provides a brief description of test data generation techniques, \Cref{subsec:PMR} gives a brief description of PMR.

\subsection{Metamorphic Testing}
\label{subsec:Metamorphic Testing}
MT is a software testing technique that aims to address the test oracle problem by using internal properties of a SUT to verify expected outputs or generate new test cases. This is achieved by analysing the relations between the inputs and outputs of multiple SUT executions, known as MRs. MRs provide descriptions of how outputs should change with specific variations in the input. The MT process typically involves five main steps:
\begin{enumerate}
    \item Create a set of initial test data (\textit{td}).
    \item Identify a list of MRs that the SUT should satisfy.
    \item Generate follow-up test data (\textit{ftd}) by applying selected MR-specified transformations to the inputs.
    \item Execute the corresponding SUT with \textit{td} and \textit{ftd}.
    \item Verify that the changes in the output observed in the \textit{td} and \textit{ftd} match the changes defined by the MR.
\end{enumerate}

The final step requires further analysis to determine the outcome of the MT workflow, as no violations do not guarantee that the SUT is implemented correctly. If an MR is violated, it suggests a fault in the SUT, assuming the MR is correctly.

\subsection{Test Data Generation}
\label{subsec:TDG}
In software testing, test data refers to the input data used during test execution. It is utilised for positive testing to verify that the functions of the SUT produce expected outputs for specific inputs and for negative testing to evaluate the SUT's ability to handle unusual or unexpected inputs. Insufficiently constructed test data could result in some potential test cases being missed, negatively impacting the software's quality. 

This paper does not aim to introduce new techniques for test data generation. Instead, we leverage Fuzz Testing for generating test data. Fuzz Testing is a software testing technique that injects erroneous or random data into software systems to discover coding errors and security vulnerabilities \cite{FuzzingSurvey}. Fuzzing consists of three main components: the input generator, executor, and defect monitor. The input generator provides the executor with various inputs. The executor runs target programs on these inputs. The defects monitors the execution to determine whether it discovers new execution states or defects \cite{FuzzingSurvey}. 

\subsection{Predicting Metamorphic Relations }
\label{subsec:PMR}
PMR was introduced by \citeauthor{PMR3}~\cite{PMR3} as an approach to predict whether a particular MR can be used to test a method in a newly developed SUT. PMR consists of three phases, regardless of whether the feature extraction mechanism is CFG-based or software metrics. In Phase I, features are extracted from the method, either by a graph description representation derived from the method's CFG or by extracting desired software metrics from the SUT source code. Phase II involves data preparation, including encoding categorical features, to meet the requirements of the ML algorithms. Finally, in Phase III, binary classification models are trained and evaluated to predict whether a specific MR is relevant to the unit testing of a particular method.

It is essential to note that the PMR approach requires a labelled dataset with a sample of methods where each method is assigned a label from a set of pre-defined MRs to use. This dataset is used to train and evaluate the binary classification models required to predict whether a specific MR is relevant to the unit testing of a particular method. Without this labelled dataset, the PMR approach cannot be applied. To evaluate the PMR approach, \citeauthor{PMR3} used a code corpus of 100 Java methods from open-source libraries such as Colt Project, Apache Mahout, Apache Commons Mathematics, Java Collections, and six pre-defined MRs. They manually labelled each method as either matching or not matching each pre-defined MR, resulting in a binary label of $1$ or $0$, respectively, to create a training dataset.
\begin{figure}[tp!]
	\centering
 	\includegraphics[width=\linewidth]{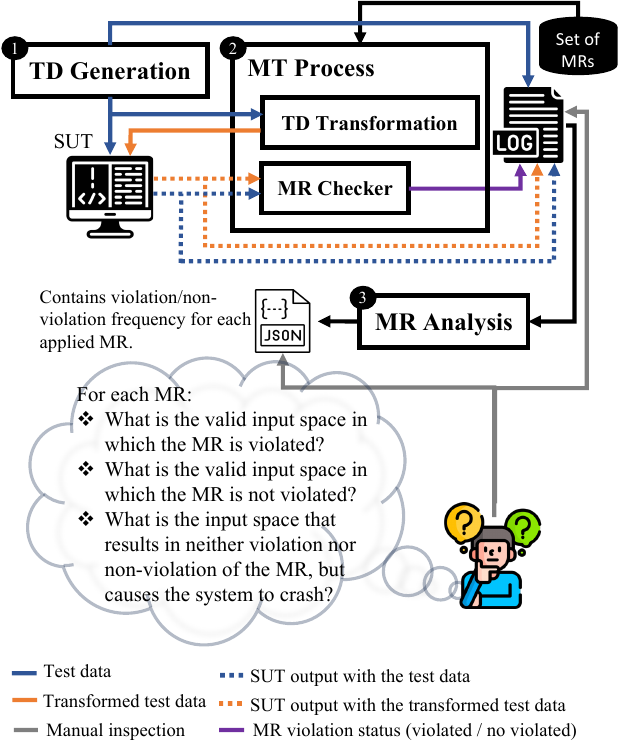}
	\caption{MetaTrimmer overview workflow}
	\label{fig:method}
\end{figure}

\section{{Methodology}}
\label{sec:methodology}
This section introduces \ad{MetaTrimmer} (\Cref{subsec:metatrimmer}) and details the experimental setup for the evaluation. This includes information about the MRs as well as the methods used as a SUT (\Cref{subsec:Metamorphic-relations}).

\subsection{MetaTrimmer}
\label{subsec:metatrimmer}

\ad{\Cref{fig:method} provides an overview of MetaTrimmer. Overall, MetaTrimmer consists of three main steps: Test Data Generation (TD Generation), MT Process, and MR Analysis. In the TD Generation step, test data (TD) is generated. The MT Process step consists of three internal processes. Firstly, the TD produced in the TD Generation step is transformed based on the MRs specifications, resulting in transformed test data (TTD). Next, both the TD and TDT are executed against the SUT, and the outputs are compared against the MR specifications using the MR Checker, which provides a violation status for each MR. Throughout these steps, the TD, TDT, corresponding SUT outputs, and the violation status of each predefined MR are stored in a log file. Finally, in the MR Analysis step, the recorded data is analysed to determine the frequency of MR violations and non-violations for each applied MR. The output of this step is a \texttt{JSON} file that contains such information.} \ad{The selection and constraint process involves manual analysis and inspection. The frequency of MR violations or non-violations is examined to determine their applicability to the SUT. A consistent 100\% violation rate indicates that the MR does not apply to the SUT, while a non-violation rate suggests a match with the SUT behaviour. In scenarios with mixed cases, \textit{i.e.,} where violations and non-violations are not 100\%, it becomes crucial to identify specific TD or ranges where the MR is applicable. This situation raises important questions for the tester: \textit{`What is the valid input space in which the MR is violated or not violated?'} and \textit{`What is the input space that results neither violates nor non-violation of MRs, but causes the system to crash?'.} By considering these questions and performing the necessary analysis, the tester can not only select MRs but also accurately specify the input-output relations across the valid input space.}

\subsection{System under test and pre-defined set of MRs}
\label{subsec:Metamorphic-relations}

As we explain in \Cref{subsec:PMR}, %\citeauthor{PMR3} used 100 Java methods obtained from open-source Java libraries. To get the ground truth data for training the models to perform the PMR approach,  
 \citeauthor{PMR3} manually labelled $100$ Java methods with a set of six pre-defined MRs in a binary manner. %This means that if a method $m$ matched a particular MR MR$_{n}$, it was labelled as $1$; if not, it was labelled as $0$. 
 In a subsequent study, \citeauthor{sanerRene}~ \cite{sanerRene} extended the original PMR approach to include two additional programming languages, Python and C++. They created two datasets consisting of source codes of methods written in Python and C++. The methods in each dataset are functionally identical to those used in the original Java dataset used by \citeauthor{PMR3}. Because the functionality of the Python methods is equivalent to that of the Java methods, \citeauthor{sanerRene} postulated that the same MRs that matched the Java methods would also match the corresponding Python and C++ methods. 
 
\begingroup
\setlength{\tabcolsep}{6pt} % Default value: 6pt
\renewcommand{\arraystretch}{1} % Default value: 1
\begin{table}[ht!]
\centering
\caption{MRs used and the total number of methods to which a specific MR applies (column `\cmark')}
{
\label{tbl:MR_specifications}
\resizebox{\linewidth}{!} {
\begin{tabular}{l|l|l|c|c}
\toprule
\textbf{MR} & \textbf{Change in the input} & \textbf{Output expected} & \cmark  & \xmark\\
\toprule
    MR$_{PER}$ & Permute the components          & Remain constant             & $23$ & $2$  \\
    MR$_{ADD}$ & Add a positive constant         & Increase or remain constant & $21$ & $4$  \\
    MR$_{MUL}$ & Multiply by a positive constant & Increase or remain constant & $24$ & $1$  \\
    MR$_{INV}$ & Take the inverse of each element& Decrease or remain constant & $20$ & $5$  \\
    MR$_{INC}$ & Add a new element               & Increase or remain constant & $14$ & $11$ \\
    MR$_{EXC}$ & Remove an element               & Decrease or remain constant & $13$ & $12$ \\
\midrule
\multicolumn{3}{r|}{\textbf{Total number of cases to which the set of MRs applies (\cmark) and does not apply (\xmark)}} & $115$ & $35$ \\
\bottomrule
\end{tabular}}}
\end{table}
\endgroup

In this paper, we used 25 Python methods from the dataset created by \citeauthor{sanerRene}. We also kept the same MRs and their labels, ensuring that they are the same as those created by \citeauthor{PMR3}\cite{PMR3}. \Cref{tbl:MR_specifications} summarises the MRs used, the changes in the inputs and expected outputs, and the total number of methods to which a specific MR applies. \Cref{tbl:MR_GT} presents the list of methods with their respective labels as reported in \cite{PMR3} and \cite{sanerRene}. The label `$1$' indicates that the MR applies (always), while the label `$0$' indicates that the MR does not apply (always) to the method. Additionally,

\begingroup
\setlength{\tabcolsep}{6pt} % Default value: 6pt
\renewcommand{\arraystretch}{1} % Default value: 1
\begin{table}[ht!]
\centering
\caption{Set of methods with labels from ~\cite{PMR3} and \cite{sanerRene}: `1' denotes the MR-method applies (always);  `0'~denotes that the MR-method does not applies (always)}
{
	\label{tbl:MR_GT}
	\resizebox{\linewidth}{!} {
	\begin{tabular}{l|c|c|c|c|c|c}
		\toprule
		\textbf{Method name} & 
            \textbf{MR$_{PER}$} & 
            \textbf{MR$_{ADD}$} &
            \textbf{MR$_{MUL}$} &
            \textbf{MR$_{INV}$} &
            \textbf{MR$_{INC}$} &
            \textbf{MR$_{EXC}$} \\
		\toprule

add\_values	        &	1	&	1	&	1	&	1	&	1	&	1	\\
average	            &	1	&	1	&	1	&	1	&	0	&	0	\\
checkNonNegative    &	1	&	0$^*$   &	1	&	1	&	1	&	0	\\
checkPositive	    &	1	&	0$^*$	&	1	&	1	&	1	&	0	\\
cnt\_zeros	        &	1	&	0$^*$	&	0$^*$	&	0$^*$	&	1	&	1	\\
count\_non\_zeros	&	1	&	1	&	1	&	1	&	1	&	1	\\
durbinWatson	    &	0	&	0	&	1	&	0	&	0	&	0	\\
entropy	            &	1	&	1	&	1	&	0	&	1	&	1	\\
find\_magnitude	    &	1	&	1	&	1	&	1	&	1	&	1	\\
find\_max	        &	1	&	1	&	1	&	1	&	1	&	1	\\
find\_max2	        &	0	&	1	&	1	&	1	&	1	&	1	\\
find\_median	    &	1	&	1	&	1	&	1	&	0	&	0	\\
find\_min	        &	1	&	1	&	1	&	1	&	0	&	1	\\
geometric\_mean	    &	1	&	1	&	1	&	1	&	0	&	0	\\
harmonicMean	    &	1	&	1	&	1	&	1	&	0	&	0	\\
    kurtosis	    &	1	&	1	&	1	&	0	&	0	&	0	\\
max	                &	1	&	1	&	1	&	1	&	1	&	1	\\
min	                &	1	&	1	&	1	&	1	&	0	&	0	\\
product	            &	1	&	1	&	1	&	1	&	1	&	1	\\
safeNorm	        &	1	&	1	&	1	&	1	&	1	&	1	\\
sampleVariance	    &	1	&	1	&	1	&	1	&	0	&	0	\\
skew	            &	1	&	1	&	1	&	0	&	0	&	0	\\
sum	                &	1	&	1	&	1	&	1	&	1	&	1	\\
sumOfLogarithms	    &	1	&	1	&	1	&	1	&	1	&	1	\\
variance	        &	1	&	1	&	1	&	1	&	0	&	0	\\
\bottomrule
% \multicolumn{7}{l}{\textbf{$^*$} Ground Truth}\\
\end{tabular}}}
\end{table}
\endgroup

\begin{figure}[tp!]
	\centering
 	\includegraphics[ width=\linewidth]{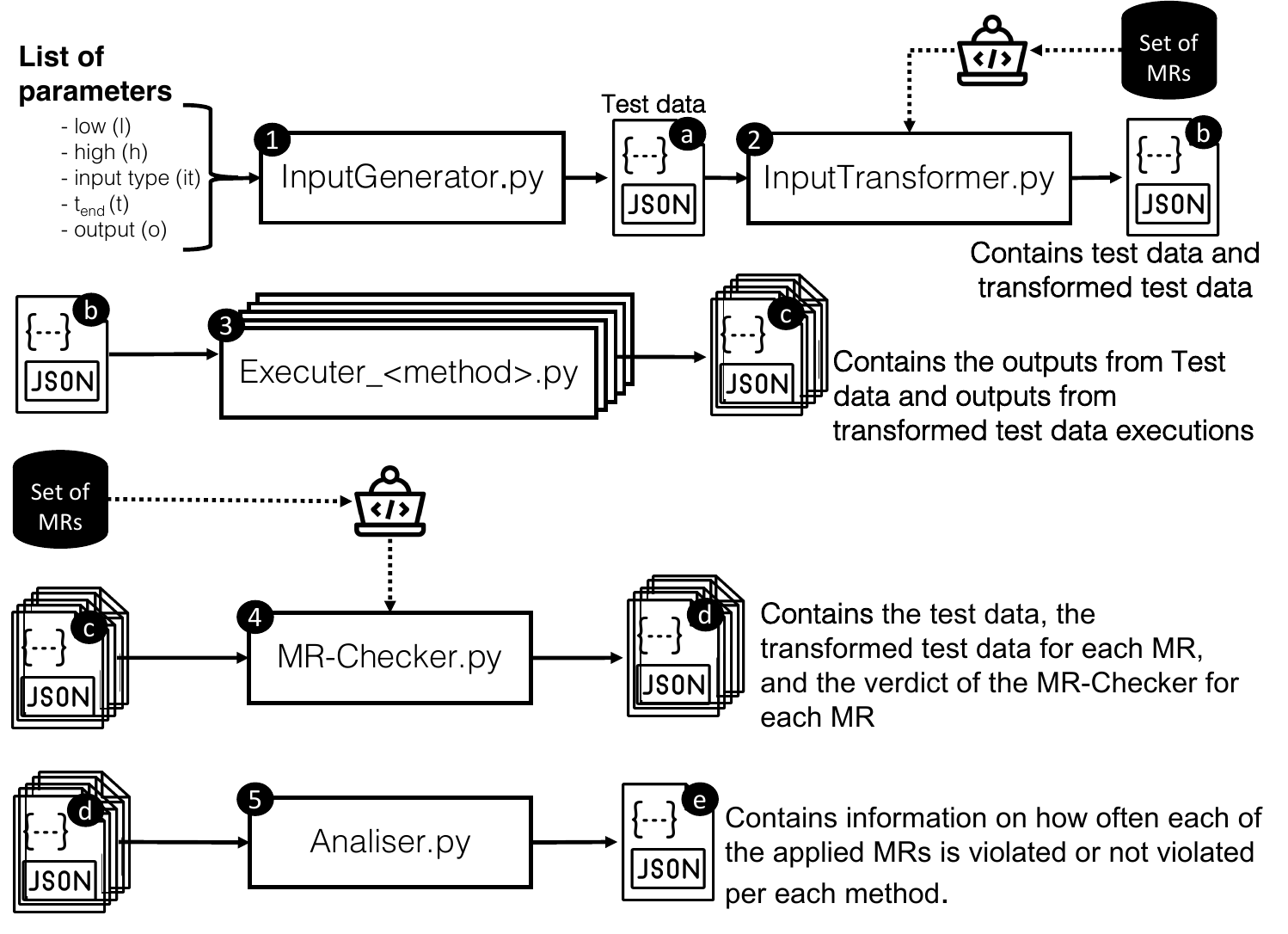}
	\caption{Possible implementation pipeline of MetaTrimmer.}
	\label{fig:pipeline}
\end{figure}

\section{{Results and Discussion}}
\label{sec:results and discussion}
\ad{A replication package with the full set of data generated during our experiments as well as all scripts 
can be found in our GitHub repo\footnote{\href{https://tinyurl.com/MetaTrimmer}{https://tinyurl.com/MetaTrimmer}}.}
\Cref{fig:pipeline} illustrates the pipeline that was developed for implementing MetaTrimmer and conducting the evaluations. The implementation consists of five Python scripts. The first script, named \texttt{InputGenerator.py}\tikz[anchor=base,baseline]\pic{mycirc=1};, is a fuzzer that uses a random number generator to create the test data, which is based on a list of parameters. The output of this script is a \texttt{JSON}\tikz[anchor=base,baseline]\pic{mycirc=a}; file that contains the generated test data. The parameters are:

\begin{itemize}
    \item low \texttt{(l)}: Fuzzer minimum value threshold.
    %The minimum number that the fuzzer will consider. %For instance, if ``low" is set to -1, the fuzzer will not generate elements less than -1.
\item high \texttt{(h)}: Fuzzer maximum value threshold. %The maximum number the fuzzer will consider. %For instance, if ``high" is set to 5, the fuzzer will not generate elements greater than 5.
\item input\_type \texttt{(it)}: The type of elements in the list. Currently, the only available options are `int' or `float'.
\item $t_{end}$ \texttt{(t)}: The end time of the program execution.
\item output \texttt{(o)}: The name of the output file.

\end{itemize}

The second script,  \texttt{InputTransformer.py}\tikz[anchor=base,baseline]\pic{mycirc=2};, transforms the test data generated by \texttt{InputGenerator.py} based on the rules defined by each MR. The \texttt{InputTransformer.py} script takes the \texttt{JSON}\tikz[anchor=base,baseline]\pic{mycirc=a}; file containing the generated test data as input and produces a new \texttt{JSON}\tikz[anchor=base,baseline]\pic{mycirc=b}; file with the transformed data as output. The number of fields in the \texttt{JSON}\tikz[anchor=base,baseline]\pic{mycirc=b}; file will depend on the number of MRs used. For instance, we used six MRs, so we expected to have seven fields: the original test data and the transformed data for each of the six MRs. It is important to note that the MRs are described in natural language and need to be translated into code so that they can be applied to the test data. The MRs are hardcoded in \texttt{InputTransformer.py} script so that they are applied automatically to the generated test data to generate the transformed test data.

As \Cref{fig:pipeline} shows, for each method tested, there exists an \texttt{Executer\_<method>.py}\tikz[anchor=base,baseline]\pic{mycirc=3}; script. The main purpose of the executer scripts is to run the original test data, and the transformed test data generated for each MR, \textit{i.e., }\texttt{JSON}\tikz[anchor=base,baseline]\pic{mycirc=b};, and store their outputs, represented by output \texttt{JSON}\tikz[anchor=base,baseline]\pic{mycirc=d}; in the \Cref{fig:pipeline}. The number of executers required depends on the number of methods being tested. In our experiment, we tested 25 different methods, creating 25 different executers and getting 25 \texttt{JSON}\tikz[anchor=base,baseline]\pic{mycirc=c}; files. 

Then, the \texttt{MR-Checker.py}\tikz[anchor=base,baseline]\pic{mycirc=4}; script consumes the outputs generated by the executer scripts and checks whether the outputs of the original test data and the transformed test data match the corresponding MR. Similar to \texttt{InputTransformer.py} script, the expected output relations for each MR are hardcoded into the \texttt{MR-Checker.py} script, which means that the script applies these relations automatically to the generated outputs from the executer scripts. After the test data is generated, transformed, and checked, a \texttt{JSON}\tikz[anchor=base,baseline]\pic{mycirc=d}; file is produced per each method containing information about the execution ID, the original test data, the transformed test data for each MR, and the verdict of the MR-Checker for each MR. 
The \texttt{JSON}\tikz[anchor=base,baseline]\pic{mycirc=d}; files produced by the MR-Checker are used as input for the final script, called \texttt{Analiser.py}\tikz[anchor=base,baseline]\pic{mycirc=5};. The \texttt{Analiser.py} script's purpose is to calculate for each method how often each of the applied MRs is violated or not violated, and store this information, \texttt{JSON}\tikz[anchor=base,baseline]\pic{mycirc=e}; in \Cref{fig:pipeline}.

\begin{figure}[ht!]
	\centering
 	\includegraphics[trim= 6mm 1mm 15mm 11mm, width=\linewidth]{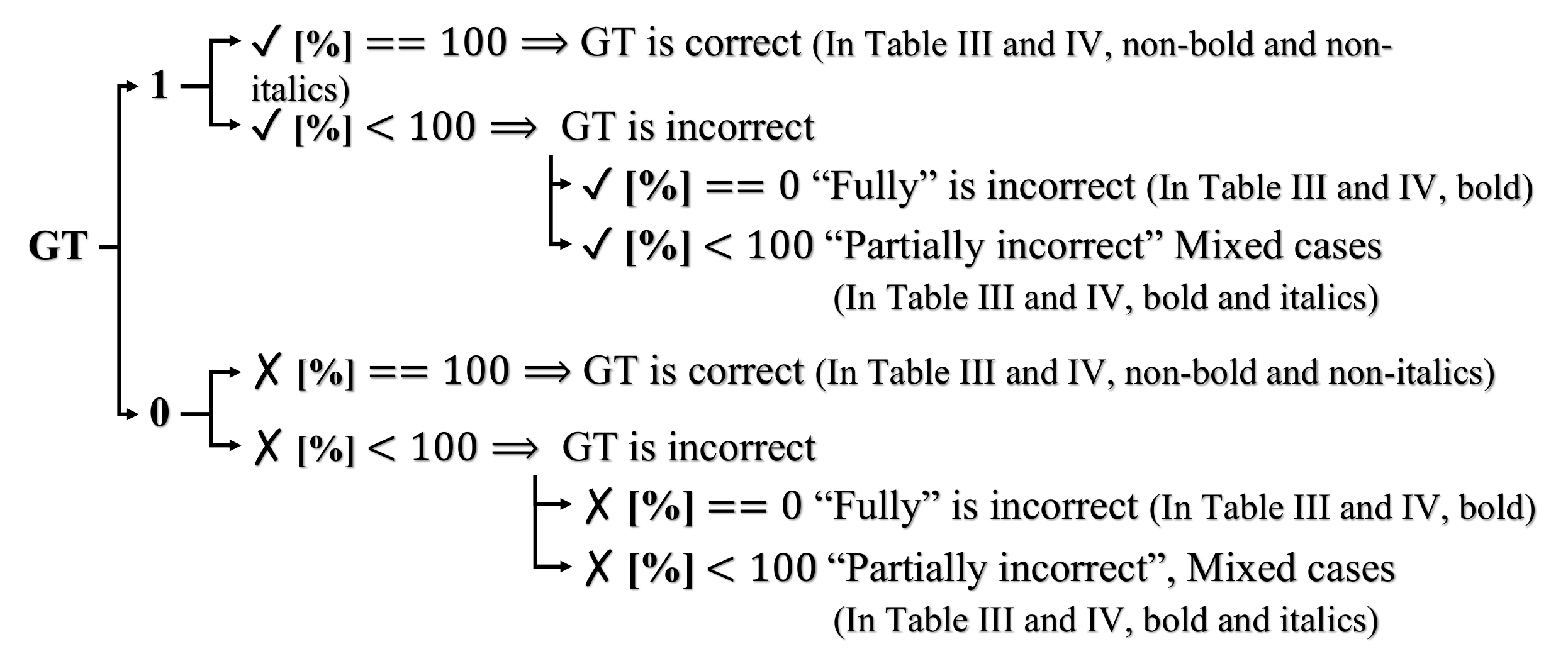}
	\caption{ 
 GT values for \ad{MetaTrimmer} evaluation: 1 means the MR always applies, while 0 means it doesn't always apply. We use symbols \cmark and \xmark~to indicate the percentage of runs where the MR applies or is violated. \cmark at 100\% means the GT is correct. Otherwise, it's incorrect. If \cmark~is 0, GT is fully incorrect; if it's less than 100\% but not 0, it's partially incorrect (mixed case). Similarly, \xmark~ at 100\% means the GT is correct. Otherwise, it could be partially incorrect (mixed case), but if \xmark~ is 0, the GT is incorrect.}
 \label{fig:gt}
\end{figure}

% \vspace{-5ex}

\subsection{RQ$_{1}$:~How well does the proposed method perform compared to the PMR approach in terms of MR selection?} 
\label{subsec:results_RQ1}

In RQ$_{1}$, we are interested to know whether an MR applies to a specific method. We hypothesise that if the MR is violated in all test data, it does not apply to the tested method. Conversely, if the MR is not violated for all test data, it applies to the tested method. \ad{MetaTrimmer} differs from the PMR method in that it chooses MRs based on test data, while PMR employs binary classifiers to predict if a given method consistently applies to a specific MR. Thus, comparing the performance of the proposed method with the PMR approach in terms of MR selection is not appropriate. However, we can compare the outcomes of \ad{MetaTrimmer} with the PMR approach in terms of the number of MRs that apply to a given method based on its Ground Truth (GT). We assumed that the GT labels were intended for only positive numbers. As such, we generated test data with the constraint of using positive numbers only, and we avoided any possible exceptions that may arise due to invalid inputs. To generate the test data, we used \texttt{InputTransformer.py} script with the parameters specified below: \texttt{l}$=1$, \texttt{h}$=50$, \texttt{it}$=$ int, \texttt{t}$=0.5$ s.

\begingroup
\setlength{\tabcolsep}{6pt} % Default value: 6pt
\renewcommand{\arraystretch}{1.2} % Default value: \cmark
\begin{table*}[ht!]
\begin{spacing}{1}
\centering
\caption{Set of methods with the GT from ~\cite{PMR3} and \cite{sanerRene}: `1' means that the MR always applies, and `0' means that the MR does not always apply. Symbol~\cmark~denotes the percentage of runs when the MR applies, and symbol~\xmark~denotes the percentage of runs when the MR does not apply (= is violated). Test data restriction: only positive integer numbers}
{
	\label{tbl:RQ1}
	\resizebox{\linewidth}{!} {
	\begin{tabular}{l|ccc|ccc|ccc|ccc|ccc|ccc}
		\toprule

      \multirow{2}{*}{\textbf{Method name}}
    & \multicolumn{3}{c|} {\textbf{\textbf{MR$_{PER}$}}}
    & \multicolumn{3}{c|} {\textbf{\textbf{MR$_{ADD}$}}}
    & \multicolumn{3}{c|} {\textbf{\textbf{MR$_{MUL}$}}}
    & \multicolumn{3}{c|} {\textbf{\textbf{MR$_{INV}$}}}
    & \multicolumn{3}{c|} {\textbf{\textbf{MR$_{INC}$}}}
    & \multicolumn{3}{c} {\textbf{\textbf{MR$_{EXC}$}}}\\
            &
            \textbf{GT} & \cmark~[\%] & \xmark~[\%] &
            \textbf{GT} & \cmark~[\%] & \xmark~[\%] &
            \textbf{GT} & \cmark~[\%] & \xmark~[\%] &
            \textbf{GT} & \cmark~[\%] & \xmark~[\%] &
            \textbf{GT} & \cmark~[\%] & \xmark~[\%] &
            \textbf{GT} & \cmark~[\%] & \xmark~[\%] \\
		\toprule

add\_values	        &	1	&	100	&	0	&	1	&	100	&	0	&	1	&	100	&	0	&	1	&	100	&	0	&	1	&	100	&	0	&	1	&	100	&	0	\\
average	            &	1	&	100	&	0	&	1	&	100	&	0	&	1	&	100	&	0	&	1	&	100	&	0	&	0	&	0	&	100	&	\textit{0}	&	\textit{50}	&	\textit{50}	\\
checkNonNegative	&	1	&	100	&	0	&	\textbf{0}	&	\textbf{100}	&	\textbf{0}	&	1	&	100	&	0	&	1	&	100	&	0	&	1	&	100	&	0	&	\textbf{0}	&	\textbf{100}	&	\textbf{0}	\\
checkPositive	    &	1	&	100	&	0	&	\textbf{0}	&	\textbf{100}	&	\textbf{0}	&	1	&	100	&	0	&	1	&	100	&	0	&	1	&	100	&	0	&	\textbf{0}	&	\textbf{100}	&	\textbf{0}	\\
cnt\_zeros	        &	1	&	100	&	0	&	\textbf{0}	&	\textbf{100}	&	\textbf{0}	&	\textbf{0}	&	\textbf{100}	&	\textbf{0}	&	\textbf{0}	&	\textbf{100}	&	\textbf{0}	&	1	&	100	&	0	&	1	&	100	&	0	\\
count\_non\_zeros   &	1	&	100	&	0	&	1	&	100	&	0	&	1	&	100	&	0	&	1	&	100	&	0	&	1	&	100	&	0	&	1	&	100	&	0	\\
durbinWatson	    &	\textit{0}	&	\textit{4}	&	\textit{96}	&	0	&	0	&	100	&	1	&	100	&	0	&	\textit{0}	&	\textit{14}	&	\textit{86}	&	\textit{0}	&	\textit{88}	&	\textit{12}	&	\textit{0}	&	\textit{48}	&	\textit{52}	\\
entropy	            &	1	&	100	&	0	&	1	&	100	&	0	&	1	&	100	&	0	&	\textit{0}	&	\textit{89}	&	\textit{11}	&	1	&	100	&	0	&	1	&	100	&	0	\\
find\_magnitude	    &	1	&	100	&	0	&	1	&	100	&	0	&	1	&	100	&	0	&	1	&	100	&	0	&	1	&	100	&	0	&	1	&	100	&	0	\\
find\_max	        &	1	&	100	&	0	&	1	&	100	&	0	&	1	&	100	&	0	&	1	&	100	&	0	&	1	&	100	&	0	&	1	&	100	&	0	\\
find\_max2	        &	\textit{0}	&	\textit{14}	&	\textit{86}	&	1	&	100	&	0	&	1	&	100	&	0	&	1	&	100	&	0	&	1	&	100	&	0	&	1	&	100	&	0	\\
find\_median	    &	1	&	100	&	0	&	1	&	100	&	0	&	1	&	100	&	0	&	1	&	100	&	0	&	\textit{0}	&	\textit{9}	&	\textit{91}	&	\textit{0}	&	\textit{54}	&	\textit{46}	\\
find\_min	        &	1	&	100	&	0	&	1	&	100	&	0	&	1	&	100	&	0	&	1	&	100	&	0	&	\textit{0}	&	\textit{49}	&	\textit{51}	&	\textbf{1}	&	\textit{86}	&	\textit{14}	\\
geometric\_mean	    &	1	&	100	&	0	&	1	&	100	&	0	&	1	&	100	&	0	&	1	&	100	&	0	&	0	&	0	&	100	&	\textit{0}	&	\textit{59}	&	\textit{41}	\\
harmonicMean	    &	1	&	100	&	0	&	1	&	100	&	0	&	1	&	100	&	0	&	1	&	100	&	0	&	\textit{0}	&	\textit{3}	&	\textit{97}	&	\textit{0}	&	\textit{69}	&	\textit{31}	\\
kurtosis	        &	1	&	100	&	0	&	1	&	100	&	0	&	1	&	100	&	0	&	\textit{0}	&	\textit{17}	&	\textit{83}	&	\textit{0}	&	\textit{48}	&	\textit{52}	&	\textit{0}	&	\textit{65}	&	\textit{35}	\\
max	                &	1	&	100	&	0	&	1	&	100	&	0	&	1	&	100	&	0	&	1	&	100	&	0	&	1	&	100	&	0	&	1	&	100	&	0	\\
min	                &	1	&	100	&	0	&	1	&	100	&	0	&	1	&	100	&	0	&	1	&	100	&	0	&	\textit{0}	&	\textit{49}	&	\textit{51}	&	\textit{0}	&	\textit{86}	&	\textit{14}	\\
product	            &	1	&	100	&	0	&	1	&	100	&	0	&	1	&	100	&	0	&	1	&	100	&	0	&	1	&	100	&	0	&	1	&	100	&	0	\\
safeNorm	        &	1	&	100	&	0	&	1	&	100	&	0	&	1	&	100	&	0	&	1	&	100	&	0	&	1	&	100	&	0	&	1	&	100	&	0	\\
sampleVariance	    &	1	&	100	&	0	&	1	&	100	&	0	&	1	&	100	&	0	&	1	&	100	&	0	&	0	&	0	&	100	&	0	&	0	&	100	\\
skew	            &	1	&	100	&	0	&	1	&	100	&	0	&	1	&	100	&	0	&	\textit{0}	&	\textit{11}	&	\textit{89}	&	\textit{0}	&	\textit{81}	&	\textit{19}	&	\textit{0}	&	\textit{50}	&	\textit{50}	\\
sum	                &	1	&	100	&	0	&	1	&	100	&	0	&	1	&	100	&	0	&	1	&	100	&	0	&	1	&	100	&	0	&	1	&	100	&	0	\\
sumOfLogarithms	    &	1	&	100	&	0	&	1	&	100	&	0	&	1	&	100	&	0	&	1	&	100	&	0	&	1	&	100	&	0	&	1	&	100	&	0	\\
variance	        &	1	&	100	&	0	&	1	&	100	&	0	&	1	&	100	&	0	&	1	&	100	&	0	&	\textit{0}	&	\textit{91}	&	\textit{9}	&	\textit{0}	&	\textit{46}	&	\textit{54}	\\
\bottomrule
\multicolumn{7}{l}{\textbf{GT:} Ground Truth}\\
\end{tabular}}}
\end{spacing}
\end{table*}
\endgroup

\Cref{fig:gt} shows the possible values of GT and its corresponding meaning for \ad{MetaTrimmer}. When GT is set to `$1$', the MR always applies. On the other hand, a value of `$0$' means that the MR does not always apply. The symbol~\cmark~denotes the percentage of runs when the MR applies, and it has two possible outcomes: If \cmark is 100\%, we assume that the GT is correct. These cases are marked in \Cref{tbl:RQ1} and \Cref{tbl:RQ2} using non-bold and non-italics formatting. However, if it is less than 100\%, we assume the GT is incorrect.  In such cases, we distinguish between two types of incorrectness: if \cmark is equal to $0$, it means that the GT is fully incorrect. These cases are marked in \Cref{tbl:RQ1} and \Cref{tbl:RQ2}. If \cmark is less than 100\% but not equal to 0, we refer to it as a mixed case, where the GT is partially incorrect. These cases are marked in \Cref{tbl:RQ1} and \Cref{tbl:RQ2} using bold and italics formatting. Similarly, the symbol~\xmark~denotes the percentage of runs where the MR does not apply or is violated. If \xmark is 100\%, we assume that the GT is correct. These cases are marked in \Cref{tbl:RQ1} and \Cref{tbl:RQ2} using non-bold and non-italics formatting. However, if it is less than 100\%, we may consider it a mixed case, where the GT could be partially incorrect. These cases are marked in \Cref{tbl:RQ1} and \Cref{tbl:RQ2} using italics formatting. If \xmark is equal to 0, it means that the GT is incorrect. These cases are marked in \Cref{tbl:RQ1} and \Cref{tbl:RQ2} using bold formatting.

\Cref{tbl:RQ1} shows the methods used, their GT, and the frequency with which each applied MR is violated or not violated per method. While analysing \Cref{tbl:RQ1}, at least four distinct patterns can be drawn: \textit{i)} Methods labelled as $1$ in the GT showed a 100\% compliance rate, indicating that the MRs may apply to those methods. \textit{ii)} In only 5 out of 35 cases, an MR was violated 100\% of the time. This is for, \texttt{durbinWatson} violated MR$_{ADD}$ 100\% of the time, while \texttt{average}, \texttt{geometric\_mean}, and \texttt{sampleVariance} violated MR$_{INC}$ and \texttt{sampleVariance} violated MR$_{EXC}$, all 100\% of the time. \textit{iii)} Some methods had a small proportion of non-violations (less than 17\%) and a high proportion of violations. This was observed in 7 out of 35 cases. For instance, \texttt{durbinWatson} violated MR$_{PER}$ only 4\% of the time, while \texttt{harmonic\_mean} violated it only 3\% of the time. 

These findings suggest that there is not always a scenario where an MR cannot apply to a specific method. However, it is important to note that such findings do not render the MRs useless. Instead, investigating and understanding the small proportion of non-violations can be valuable in improving the MRs. By making them more precise and customised to the specific test data. \textit{iv)} the analysis revealed cases where violations and non-violations were relatively proportional. This was observed in 11 out of 34 cases, where the violations and non-violations were proportional, indicating that the MRs but with exceptions. For instance, \texttt{kurtosis} violated MR$_{INC}$ 52\% of the time and had no violations 48\% of the time, while for MR$_{EXC}$, it had 35\% violations and 65\% non-violations. This implies that exceptions must be considered and accounted for when applying MRs. In such cases, defining constraints based on the specific test data is even more important.

Regarding $RQ_1$, our findings show that \ad{MetaTrimmer} for selecting MRs performs better than the PMR approach, with a compliance rate of 100\% for all methods labelled as ‘1’ in the GT. Furthermore, we identified three methods with incorrect labels. Additionally, \ad{MetaTrimmer} identified cases where the MRs do not apply, cases with a small proportion of non-violations and a high proportion of violations, as well as cases where violations and non-violations are relatively proportional for methods labelled as ‘0’ in the GT. These findings suggest that there is not always a scenario where an MR cannot apply to a specific method. By investigating and understanding these non-violations, we can improve the MRs by making them more precise and customised to the specific test data, \textit{i.e.,} adding constraints. Even with initial constraints for the test data (only positive integer numbers), \ad{MetaTrimmer} can provide valuable constraints for MRs that may only apply to specific test data. These constraints can help uncover situations that may improve the overall test coverage of the test suite.

\begingroup
\setlength{\tabcolsep}{6pt} % Default value: 6pt
\renewcommand{\arraystretch}{1.2} % Default value: \cmark
\begin{table*}[ht!]
\begin{spacing}{1}
\centering
\caption{Set of methods with the GT from ~\cite{PMR3} and \cite{sanerRene}: `1' means that the MR always applies, and `0' means that the MR does not always apply. Symbol~\cmark~denotes the percentage of runs when the MR applies, and symbol~\xmark~denotes the percentage of runs when the MR does not apply. Test data restriction: only integers. Numbers in parentheses refer to the percentage of invalid input data (crashes).}
{
	\label{tbl:RQ2}
	\resizebox{\linewidth}{!} {
	\begin{tabular}{l|ccc|ccc|ccc|ccc|ccc|ccc}
		\toprule

      \multirow{2}{*}{\textbf{Method name}}
    & \multicolumn{3}{c} {\textbf{\textbf{MR$_{PER}$}}}
    & \multicolumn{3}{c} {\textbf{\textbf{MR$_{ADD}$}}}
    & \multicolumn{3}{c} {\textbf{\textbf{MR$_{MUL}$}}}
    & \multicolumn{3}{c} {\textbf{\textbf{MR$_{INV}$}}}
    & \multicolumn{3}{c} {\textbf{\textbf{MR$_{INC}$}}}
    & \multicolumn{3}{c} {\textbf{\textbf{MR$_{EXC}$}}}\\
            &
            \textbf{GT} & \cmark~[\%] & \xmark~[\%] &
            \textbf{GT} & \cmark~[\%] & \xmark~[\%] &
            \textbf{GT} & \cmark~[\%] & \xmark~[\%] &
            \textbf{GT} & \cmark~[\%] & \xmark~[\%] &
            \textbf{GT} & \cmark~[\%] & \xmark~[\%] &
            \textbf{GT} & \cmark~[\%] & \xmark~[\%] \\
		\toprule

add\_values	        &
1	&	100	&	0	&	
1	&	100	&	0	&	
\textbf{\textit{1}}	& \textbf{\textit{46}} & \textbf{\textit{54}} & 
\textit{\textbf{1}}	& \textbf{\textit{37}} & \textbf{\textit{44}} \textbf{(19)} &
1	&	100	&	0	&	
\textbf{\textit{1}}	& \textbf{\textit{50}} & \textbf{\textit{50}} \\

average	            &	
1	&	100	&	0	&	
1	&	100	&	0	&	
\textbf{\textit{1}}	&	\textbf{\textit{46}}	&	\textbf{\textit{54}}	&	
\textbf{\textit{1}}	&	\textbf{\textit{37}}	&	\textbf{\textit{44} (19)}	&	
\textit{0}	&	\textit{80}	&	\textit{20}	&	
\textit{0}	&	\textit{50}	&	\textit{50}	\\

checkNonNegative	&	
1	&	100	&	0	&	
\textbf{0}	&	\textbf{100}	&	\textbf{0}	&	
1	&	100	&	0	&	
\textbf{\textit{1}}	&	\textbf{\textit{81}}	&	\textbf{\textit{0} (19)}	&	
1	&	100	&	0	&	
\textbf{0}	&	\textbf{98}	&	\textbf{2}	\\

checkPositive	    &	
1	&	100	&	0	&	
\textbf{0}	&	\textbf{100}	&	\textbf{0}	&	
1	&	100	&	0	&	
\textbf{\textit{1}}	&	\textbf{\textit{81}}	&	\textbf{\textit{0} (19)}	&	
1	&	100	&	0	&	
\textbf{0}	&	\textbf{98}	&	\textbf{2}	\\

cnt\_zeros	        
&	1	&	100	&	0	&	
\textbf{0}	&	\textbf{84}	&	\textbf{16}	&	
\textbf{0}	&	\textbf{100}&	\textbf{0}	&	
\textbf{0}	&	\textbf{81}	&	\textbf{0 (19)}	&	
1	&	100	&	0	&	
1	&	100	&	0	\\

count\_non\_zeros	&	
1	&	100	&	0	&	
\textbf{\textit{1}}	&	\textbf{\textit{84}}	&	\textbf{\textit{16}}	&	
1	&	100	&	0	&	
\textbf{\textit{1}}	&	\textbf{\textit{81}}	&	\textbf{\textit{0} (19)}	&	
1	&	100	&	0	&	
1	&	100	&	0	\\

durbinWatson	    &	
\textit{0}	&	\textit{4}	&	\textit{96}	&	
\textbf{0}	&	\textit{41}	&	\textit{59}	&	
1	&	100	&	0	&	
\textit{0}	&	\textit{39}	&	\textit{42} (19)	&	
\textit{0}	&	\textit{80}	&	\textit{20}	&	
\textit{0}	&	\textit{55}	&	\textit{45}	\\

entropy	    &	
1	&	100	&	0	&	
1	&	100	&	0	&	
1	&	100	&	0	&	
\textit{0}	&	\textit{58}	&	\textit{24} (18) &	
1	&	100	&	0	&	
1	&	100	&	0	\\

find\_magnitude	    &	
1	&	100	&	0	&	
\textbf{\textit{1}}	&	\textbf{\textit{60}}	&	\textbf{\textit{40}}	&	
1	&	100	&	0	&	
\textbf{\textit{1}}	&	\textbf{\textit{81}}	&	\textbf{\textit{0} (19)}	&	
1	&	100	&	0	&	
1	&	100	&	0	\\

find\_max	        &	
1	&	100	&	0	&	
1	&	100	&	0	&	
\textbf{\textit{1}}	&	\textbf{\textit{98}}	&	\textbf{\textit{2}}	&	
\textbf{\textit{1}}	&	\textbf{\textit{79}}	&	\textbf{\textit{2} (19)}	&	
1	&	100	&	0	&	
1	&	100	&	0	\\

find\_max2	        &	
\textit{0}	&	\textit{25}	&	\textit{75}	&	
1	&	100	&	0	&	
\textit{1}	&	\textit{92}	&	\textit{8}	&	
\textbf{\textit{1}}	&	\textbf{\textit{73}}	&	\textbf{\textit{8} (19)}	&	
1	&	100	&	0	&	
1	&	100	&	0	\\

find\_median	    &	
1	&	100	&	0	&	
1	&	100	&	0	&	
\textbf{\textit{1}}	&	\textbf{\textit{49}}	&	\textbf{\textit{51}}	&	
\textbf{\textit{1}}	&	\textbf{\textit{38}}	&	\textbf{\textit{43} (19)}	&	
\textit{0}	&	\textit{75}	&	\textit{25}	&	
\textit{0}	&	\textit{55}	&	\textit{45}	\\

find\_min	        &	
1	&	100	&	0	&	
1	&	100	&	0	&	
\textbf{\textit{1}}	&	\textbf{\textit{2}}	&	\textbf{\textit{98}}	&	
\textit{0}	&	\textit{2}	&	\textit{79 }(19)	&	
\textit{\textbf{0}}	&	\textbf{\textit{99}}	&	\textbf{\textit{1}}	&	
\textbf{\textit{1}}	&	\textbf{\textit{87}}	&	\textbf{\textit{13}}	\\

geometric\_mean	    &  	
\textbf{\textit{1}}	&	\textbf{\textit{59}}	&	\textbf{\textit{0} (41)}	&	
\textbf{\textit{1}}	&	\textbf{\textit{24}}	&	\textbf{\textit{19} (57)}&	
\textbf{\textit{1}}	&	\textbf{\textit{59}}	&	\textbf{\textit{0} (41)}	&	
\textbf{\textit{1}}	&	\textbf{\textit{40}}	&	\textbf{\textit{0} (60)}	&	
\textit{0}	&	\textit{19}	&	\textit{40} (41)&	
\textit{0}	&	\textit{28}	&	\textit{9} (63)	\\

harmonicMean	    &	
\textbf{\textit{1}}	&	\textbf{\textit{81}}	&	\textbf{\textit{0} (19)}	    &	
\textbf{\textit{1}}	&	\textbf{\textit{43}}	&	\textbf{\textit{23} (34)}	&	
\textbf{\textit{1}}	&	\textbf{\textit{40}}	&	\textbf{\textit{41} (19)}	&	
\textbf{\textit{1}}	&	\textbf{\textit{40}}	&	\textbf{\textit{41} (19)}	&	
\textit{0}	&	\textit{17}	&	\textit{64 (19)}	&	
\textit{0}	&	\textit{41}	&	\textit{40 (19)}	\\

kurtosis	        &	
1	&	100	&	0	&
1	&	100	&	0	&	
1	&	100	&	0	&	
\textit{0}	&	\textit{19}	&	\textit{62}	&	
\textit{0}	&	\textit{80}	&	\textit{20}	&	
\textit{0}	&	\textit{65}	&	\textit{35}	\\

max	                &	
1	&	100	&	0	&	
1	&	100	&	0	&	
\textbf{\textit{1}}	&	\textbf{\textit{98}}	&	\textbf{\textit{2}}	&	
\textbf{\textit{1}}	&	\textbf{\textit{79}}	&	\textbf{\textit{2} (19)}	&	
1	&	100	&	0	&	
1	&	100	&	0	\\

min	                &	
1	&	100	&	0	&	
1	&	100	&	0	&	
\textbf{\textit{1}}	&	\textbf{\textit{2}}	&	\textbf{\textit{98}}	&	
\textbf{\textit{1}}	&	\textbf{\textit{2}}	&	\textbf{\textit{79} (19)}	&	
\textit{0}	&	\textit{99}	&	\textit{1}	&	
\textit{0}	&	\textit{87}	&	\textit{13}	\\

product	            &	
1	&	100	&	0	&	
\textbf{\textit{1}}	&	\textbf{\textit{52}}	&	\textbf{\textit{48}}	&	
\textbf{\textit{1}}	&	\textbf{\textit{59}}	&	\textbf{\textit{41}}	&	
\textbf{\textit{1}}	&	\textbf{\textit{41}}	&	\textbf{\textit{40} (19)}	&	
\textbf{\textit{1}}	&	\textbf{\textit{59}}	&	\textbf{\textit{41}}		&	
\textbf{\textit{1}}	&	\textbf{\textit{59}}	&	\textbf{\textit{41}}		\\

safeNorm	        &	
1	&	100	&	0	&	
\textbf{\textit{1}}	&	\textbf{\textit{60}}	&	\textbf{\textit{40}}	&	
1	&	100	&	0	&	
\textbf{\textit{1}}	&	\textbf{\textit{81}}	&	\textbf{\textit{0} (19)}	&
1	&	100	&	0	&	
1	&	100	&	0	\\

sampleVariance	    &	
1	&	100	&	0	&	
1	&	100	&	0	&	
1	&	100	&	0	&	
\textbf{\textit{1}}	&	\textbf{\textit{81}}	&	\textbf{\textit{0} (19)}	&	
0	&	0	&	100	&	
0	&	0	&	100	\\

skew	            &	
1	&	100	&	0	&	
1	&	100	&	0	&	
1	&	100	&	0	&	
\textit{0}	&	\textit{40}	&	\textit{41 (19)}	&	
\textit{0}	&	\textit{18}	&	\textit{82}	&	
\textit{0}	&	\textit{50}	&	\textit{50}	\\

sum	                &	
1	&	100	&	0	&	
1	&	100	&	0	&	
\textbf{\textit{1}}	&	\textbf{\textit{46}}	&	\textbf{\textit{54}}	&	
\textbf{\textit{1}}	&	\textbf{\textit{37}}	&	\textbf{\textit{44} (19)}	&	
1	&	100	&	0	&	
\textbf{\textit{1}}	&	\textbf{\textit{50}}	&	\textbf{\textit{50}}	\\

sumOfLogarithms	    &	
\textbf{\textit{1}}	&	\textbf{\textit{81}}	&	\textbf{\textit{0} (19)}	&	
\textbf{\textit{1}}	&	\textbf{\textit{34}}	&	\textbf{\textit{32} (34)}&
\textbf{\textit{1}}	&	\textbf{\textit{81}}	&	\textbf{\textit{0} (19)}	&	
\textbf{\textit{1}}	&	\textbf{\textit{81}}	&	\textbf{\textit{0} (19)}	&	
\textbf{\textit{1}}	&	\textbf{\textit{81}}	&	\textbf{\textit{0} (19)}	&	
\textbf{\textit{1}}	&	\textbf{\textit{81}}	&	\textbf{\textit{0} (19)}	\\

variance	        &	
1	&	100	&	0	&	
1	&	100	&	0	&	
1	&	100	&	0	&	
\textbf{\textit{1}}	&	\textbf{\textit{81}}	&	\textbf{\textit{0} (19)}	&	
\textit{0}	&	\textit{9}	&	\textit{91}	&	
\textit{0}	&	\textit{46}	&	\textit{54}	\\
\bottomrule
\multicolumn{15}{l}{\textbf{GT:} Ground Truth (NB: We use the same GT as in TABLE III to make the direct comparison of entries easier)}\\
\end{tabular}}}
\end{spacing}
\end{table*}
\endgroup

\subsection{RQ$_{2}$: How can test data be used to obtain constraints for MRs?}
\label{subsec:results_RQ2}

% \textit{\textbf{RQ$_{2}$: How can test data be used to obtain constraints for MRs?}} 

In RQ$_{1}$, we assumed a scenario where we had prior knowledge of the initial restriction for the test data, \textit{i.e.,} only positive numbers. In RQ$_{2}$, we investigate a scenario where we lack this prior knowledge. Thus, we aim to determine if mixed cases can provide valuable information for constraining MRs. To generate the test data, we used \texttt{InputTransformer.py} script with the parameters specified below: \texttt{l}$=-15$, \texttt{h}$=15$, \texttt{it}$=$ int, \texttt{t}$=0.5$ s.
% 
% \begin{itemize}
%     \item low \texttt{(l)}: -15
%     \item high \texttt{(h)}: 15
%     \item input\_type \texttt{(it)}: int
%     \item $t_{end}$ \texttt{(t)}: 0.5 s
% \end{itemize}
% 
 Upon initial examination of the results presented in \Cref{tbl:RQ2}, one can observe three types of outcomes: violation, non-violation, and invalid data. The percentage of invalid data is denoted in parentheses in the column that indicates that the MR does not apply (\textit{i.e., }column \xmark~[\%]). Upon analysing the results presented in \Cref{tbl:RQ2}, one can identify at least five patterns. \textit{i)} It is evident that the applicability of MRs can be easily determined by checking if there are 100\% no violations. When the GT is labelled as `1', we observe that some previously identified MRs still hold. For example, 20 out of the 23 cases in MR$_{PER}$ remain unchanged.

\textit{ii)} Unlike in \Cref{tbl:RQ1}, some GTs marked as `1' in \Cref{tbl:RQ2} have mixed cases. For instance, the \texttt{add\_values} method under MR$_{MUL}$ has 46 non-violations and 54 violations. \textit{iii)} When there is no initial constraint, a third type of outcome could happen, invalid data. For instance, the \texttt{geometric\_mean} method has only 59\% non-violations and 0\% violations, indicating that 41\% of the test data was invalid. \textit{iv)} Some methods had a small proportion of non-violations and a high proportion of violations and vice versa. \textit{v)}, similar to RQ$_1$, there are cases where the violations and non-violations were proportional. To extract constraints, we manually inspected the mixed cases for each method. This detailed examination allowed us to identify a clear pattern: most violations occurred when negative numbers were involved in the test data. We also noticed that empty arrays in the test data could result in either violations or invalid data. 

Regarding \textit{RQ$_{2}$}, our findings indicate that test data can be used to derive constraints for MRs by examining their violations and non-violations for various inputs. When an MR consistently holds for specific test data, it can be considered a constraint for the MR. Conversely, if an MR is consistently violated for specific test data, it is not appropriate for the SUT. Mixed cases may also arise, indicating specific test data or test data ranges for which the MR is applicable and providing constraints to restrict the scope of the MR accordingly. By analysing the reasons behind these mixed cases, specific constraints for the MR can be identified. We recognise that deriving constraints for MRs through manual inspection of violation and non-violation results can be a tedious and time-consuming task. To address this issue, in the next stage of our research, we plan to focus on developing data mining techniques to automate the derivation of constraints for MRs. This will involve identifying relevant features and patterns in the data to improve the analysis coverage and identify potential constraints that may have been missed during manual inspections.

% \vspace{-0.8ex}

\subsection{Threats to Validity}
\label{subsec:threats_validity}

In the context of our study, two types of threats to validity are most relevant: threats to internal and external validity.

To achieve internal validity, we used the same set of methods and MRs as in \citeauthor{PMR3}. There is a potential risk of bias in selecting the MRs or the test data used, which could potentially impact the results obtained. For instance, we found cases where certain methods and MRs were incorrectly labelled during our analysis.

To enhance external validity and ensure the generalisability of the results, it would have been preferable to include a wider range of methods in the evaluation process. This would have helped to mitigate any potential bias that could have arisen from the selection of the methods used. Therefore, the current study may not be able to determine the full extent of the effectiveness of \ad{MetaTrimmer}.

\ad{\subsection{Remarks on General Relevance}
The results obtained in our preliminary evaluation demonstrate a small yet promising potential of MetaTrimmer. It is important to note that MetaTrimmer is designed to automate the process of selecting and constraining MRs based on test data, which offers significant time and effort savings compared to manual selection and analysis. Additionally, MetaTrimmer's methodology is domain-agnostic, allowing its application to diverse software systems and domains. This practical relevance makes it a versatile solution suitable for various testing scenarios, regardless of the source code or specific domain. Furthermore, MetaTrimmer goes beyond MR selection by deriving constraints for test data. This practical feature enables the identification of specific test data or ranges where MRs are applicable, providing valuable guidance to testers in creating focused test cases and ensuring comprehensive testing coverage.} \ad{While we acknowledge the need for further validation and verification, we have plans to expand the evaluation of MetaTrimmer across multiple domains. By testing various software systems representing different domains, we aim to assess the method's effectiveness and generalisability in diverse contexts. Additionally, collaboration with other researchers and testing practitioners will be sought to validate the MetaTrimmer method in their respective environments. Sharing the approach, guidelines, and codebase will encourage adoption and facilitate gathering feedback from the community. To establish the practical relevance of MetaTrimmer, comparative studies will be conducted by applying MetaTrimmer alongside existing MR selection methods or alternative testing approaches. This comparative analysis will enable the comparison of results, efficiency, and effectiveness of MetaTrimmer against other methods, highlighting its advantages and practical relevance in the field. }
\vspace{-2ex}
\section{{Related work}}
\label{sec:related work}
Several studies have built upon the PMR approach \cite{PMR4, PMR5, vst2022Aleja, 8055540}. Despite the promising results of the PMR approach and subsequent studies, there are notable limitations. Firstly, the reliance on binary classifiers necessitates labelled datasets for effective learning. Secondly, based on CFG or source code metrics, the feature extraction process may not fully consider the effects of refactoring. Additionally, the binary output of PMR may not account for the influence of test data on MR applicability. Moreover, there is a lack of comprehensive analysis in the existing studies regarding scenarios where MRs may not always be applicable. Regarding finding constraints of MRs, \citeauthor{8785652} proposed a method that generates MRs based on two concepts: identifying potential input constraints and utilising constraint solvers to produce concrete input values satisfying the identified constraints. This approach is argued to generate more powerful and effective MRs compared to traditional methods. However, it should be noted that using constraint solving requires significant technical expertise and resources. Additionally, constraint solving can be computationally expensive and may require longer processing times, particularly for complex systems.

\section{{Conclusions}}
\label{sec:conclusion}

In this paper, we present and evaluate \ad{MetaTrimmer}, a method for selecting and constraining MRs based on test data. In RQ$_1$, we investigated whether MRs can be determined based on test data with known input restrictions. We hypothesised that if a particular MR is violated 100\%, it does not apply to the tested method. On the other hand, if the MR is not violated in 100\% of the cases, we assume that it applies to the tested method. We compared the performance of \ad{MetaTrimmer} with the PMR approach in terms of MR selection, and the results suggest that \ad{MetaTrimmer} is effective. 

In RQ$_2$, we examined whether mixed scenarios can offer insights into MR behaviour and aid in creating constraints without knowing any restrictions regarding the test data used. We hypothesised that data from mixed cases could offer insights into the behaviour of the MRs and aid in the generation of constraints. The results suggest that the method is also effective and can provide valuable information for generating constraints on MRs.
One significant outcome of our study is, indeed, that \ad{MetaTrimmer} can identify MRs that normally may have been excluded, i.e., those cases with small portions of non-violations. This can lead to a more thorough and effective test suite with improved coverage of edge cases and potential errors. Additionally, the constraints can optimise the test suite by focusing on the most critical and sensitive test data. 

It is worth noting that the constraints on MRs are derived through manual inspection of violation status, a tedious and time-consuming task. In the next stage of our research, we will focus on developing data mining techniques to identify relevant features and patterns in the data and automate the derivation of constraints. Automating this process will increase analysis coverage and identify potential constraints that may have been missed in manual inspections, allowing testers to understand the constraints and applicability of MRs better.

Finally, we would like to mention that with \ad{MetaTrimmer}, it is possible to reuse lists of known MRs as a starting point for the selection. Thus, if there existed an open-source database of MRs, which is constantly growing by adding new MRs whenever they emerge in specific contexts, then such a database could replace the tedious process of having domain experts come up with an initial set of suitable MRs to start from. 

\vspace{-2ex}

\section*{Acknowledgement}
The research reported in this paper has been partly funded by BMK, BMAW, and the State of Upper Austria in the frame of the SCCH competence center INTEGRATE [(FFG grant no. 892418)] part of the FFG COMET Competence Centers for Excellent Technologies Programme, as well as by the European Regional Development Fund, and grant PRG1226 of the Estonian Research Council.

% \balance
\printbibliography

\end{document}